\documentclass[11pt]{article}
\usepackage[]{amscd,amsfonts,amsthm}

\newtheorem{definition}{Definition}

\newtheorem{proposition}{Proposition}

\usepackage[psamsfonts]{amssymb}

\newcommand{\real}{{\mathbb R}}

\newcommand{\rational}{{\mathbb Q}}
\newcommand{\bd}{\begin{definition}}
\newcommand{\ed}{\end{definition}}
\newcommand{\bp}{\begin{proposition}}
\newcommand{\ep}{\end{proposition}}
\newcommand{\be}{\begin{equation}}
\newcommand{\ee}{\end{equation}}
\newcommand{\bea}{\begin{eqnarray}}
\newcommand{\eea}{\end{eqnarray}}
\newcommand{\ba}{\begin{array}}
\newcommand{\ea}{\end{array}}
\begin{document}

\title{Reality and Probability: Introducing a New Type of Probability Calculus\footnote{Published as: D. Aerts, ``Reality and
probability: introducing a new type of probability calculus", in {\it Probing the Structure of
Quantum Mechanics: Nonlocality, Computation and Axiomatics}, eds. D. Aerts, M. Czachor and T. Durt, World Scientific, Singapore
(2002).}}

\author{Diederik Aerts}

\date{}
\maketitle

\centerline{Center Leo Apostel (CLEA) and}
\centerline{Foundations of the Exact Sciences
(FUND),}
\centerline{Brussels Free
University, Krijgskundestraat 33,}
\centerline{1160 Brussels,
Belgium.}
\centerline{diraerts@vub.ac.be}

\begin{abstract}
\noindent
We consider a conception of reality that is the following: An object is `real' if we know that
if we would try to test whether this object is present, this test would give us the answer `yes' with certainty. The knowledge
about this certainty we gather from our overall experience with the world. If we consider a conception of reality where
probability plays a fundamental role, which we should do if we want to incorporate the microworld into our reality, it can be
shown that standard probability theory is not well suited to substitute `certainty' by means of `probability equal to 1'. We
analyze the different problems that arise when one tries to push standard probability to deliver a conception of reality as the
one we advocate. The analysis of these problems lead us to propose a new type of probability theory that is a generalization of
standard probability theory. This new type of probability theory is a function to the set of all subsets of the interval $[0,
1]$ instead of to the interval $[0, 1]$ itself, and hence its evaluation happens by means of a subset instead of a number. This
subset corresponds to the different limits of sequences of relative frequency that can arise when an intrinsic lack of knowledge
about the context and how it influences the state of the physical entity under study in the process of experimentation is taken
into account. The new probability theory makes it possible to define probability on the whole set of experiments within the
Geneva-Brussels approach to quantum mechanics, which was not possible with standard probability theory. We introduce the formal
mathematical structure of a `state experiment probability system', by using this new type of probability theory, as a general
description of a physical entity by means of its states, experiments and probability. We derive the state property system as a
special case of this structure, when we only consider the `certain' aspects of the world. The category {\bf SEP} of state
experiment probability systems and their morphisms is linked with the category {\bf SP} of state property systems and their
morphisms, that has been studied in earlier articles in detail.
\end{abstract}

\section{Introducing the Problem}
We first introduce the conceptual tools that we consider for our problem. We suppose that we have at our disposal a physical
entity $S$ that can be in different states $p, q, r, \ldots \in \Sigma$, where $\Sigma$ is the set of all relevant states of the
entity $S$. We also have at our disposal different experiments $\alpha, \beta, \gamma, \ldots \in Q$ that can be performed on
the entity $S$. We suppose that each experiments has only two possible outcomes, that we label `yes' and `no'. $Q$ is the set of
all relevant yes/no-experiments.

\subsection{What is Real} \label{subsec:real}
Let us consider an example. The physical entity $S$ that we consider is a piece of wood\footnote{This example was first introduced
in \cite{Aerts1981,Aerts1982}.}. We consider two yes/no-experiments $\alpha$ and $\beta$ that can be performed on this piece of
wood and that test respectively whether the piece of wood `burns well' and whether the piece of wood `floats on water'. The
experiment
$\alpha$ consists, for example, in setting the piece of wood on fire and observing whether it keeps burning for a sufficiently
long time. We give the outcome `yes' if this is the case, and the outcome `no' if this is not the case. The experiment $\beta$
consists of putting the piece of wood in water and observing whether it floats. We give the outcome `yes' if this is the case and
the outcome `no' if this is not the case\footnote{We remark that are types of wood that do not float on water. For example the
African wood called `wenge'.}.

We can say that experiment $\alpha$ tests the property of `burning well' of the piece of wood, let us call this property $a$, while
experiment $\beta$ tests the property of `floating on water' of the piece of wood, and we call this property $b$.

First of all we have to understand clearly when we say that a piece of wood `has' a certain property, or, in other words, when a
certain property is `actual' for a piece of wood. Let us focus on property $a$ to analyze this. We say that a
piece of wood `has' the property of `burning well' if its state is such that if we would perform experiment $\alpha$ the outcome
`yes' would come out with certainty. We remark that this is not related to the fact that we have performed the
experiment, because indeed after we have put the piece of wood on fire, and after it has burned, it will not have the property of
`burning well' any longer. The experiment $\alpha$ destroys the property $a$. This is common place in the world. However not always
an experiment that tests a property also destroys this property. This is for example not the case for property $b$. If
the state of the piece of wood was such that we could predict with certainty that it would float on water in the eventuality of
performing the experiment $\beta$, even after having performed experiment $\beta$, the state of the piece of wood will still be such.
The important thing to notice however is that this difference, between the effect of the experiments
$\alpha$ and $\beta$, the first destroying the property that it tests, and the second not doing so, does not play a role in the
aspect of the experiments that defines the actuality of the properties that they test. If we say that a Volvo is a `strong' car,
this refers to experiments that have been carried out in the Volvo factory, consisting of hitting a brick wall with the car and
measuring the amount of damage. A new Volvo however, the one of which we say that it is a strong car, has not undergone these
experiments, exactly because the experiments will destroy the car. Still we all believe that this new Volvo `has' the property of
being a strong car. We all demand that the experiment to test the strength has not been carried out on the car that we
buy, and still we believe that this car, the one that we buy, has this property of strength, because we believe that in the
eventuality of a test, the test will give the `yes' outcome with certainty.

It is important for the rest of this article to understand this subtle way of reality. What is real
is what we believe to react in a certain way with certainty to experiments that we eventually could perform, but that in general we
do not perform, because in many cases these experiments destroy what we were considering to be real. What is real is related to
what we eventually could test, but in general do not test. This is also the way we conceive of what is real in our everyday
world, as the example of the Volvo car makes clear.

\subsection{Attributing Several Properties at Once}
When do we say that a piece of wood `has' both the properties $a$ and $b$? The subtility of the nature of `what is real' already
appears in full splendor if we analyze carefully the way in which we attribute several properties at once to an entity. At
first sight one could think that attributing the two properties $a$ and $b$ to the piece of wood has
to do with performing both experiments at once, or one after the other, or \ldots, well at least performing the two
experiments in one way or another. This is wrong. It is in fact one of the deep mistakes of classical logic, where one introduces
indeed the conjunction of propositions by means of truth tables of both proposition, as if the truth of the conjunction would be
defined by verifying the truth of both propositions. We have chosen the example of the piece of wood on purpose such that this
mistake becomes obvious. We would indeed be in great difficulties with our common sense conception of reality if attributing both
properties `burning well' and `floating on water' would have to do with performing both experiments $\alpha$ and $\beta$ in one way
or another. Putting a piece of wood on fire and at the same time making it float on water does indeed not seem to be a very
interesting enterprise in collecting knowledge about the state of the piece of wood. Performing one experiment after the other, in
whatever order, also does not seem to be very fruitful. Indeed, if we first burn the piece of wood, it does not float on water
afterwards, and if we first make it float in water, it does not burn afterwards. While we all know that we can have a very simple
piece of wood in a state such that it `has' both properties at once. Most pieces of wood indeed do have both properties at once most
of the time. How do we arrive at this belief in our daily conception of reality? Let us analyze this matter.

What we do is the following. Suppose that we have a dry piece of light wood. What we say to ourselves is that whether we would
perform experiment $\alpha$ or whether we would perform experiment $\beta$, both will give the outcome `yes' with certainty, and
that is the reason that we believe that this dry piece of light wood has both properties $a$ and $b$ at once. What we just come to
say is clear for everybody I suppose. The question is how to formalize this. Let us make an attempt. Having available the two
experiments $\alpha$ and $\beta$ we make a new experiment, that we denote $\alpha \cdot \beta$, and call the product of $\alpha$
and $\beta$, as follows: The experiment $\alpha \cdot \beta$ consists of choosing one of the experiments $\alpha$ or $\beta$, and
performing this chosen experiment, and collecting the outcome, `yes' or `no', and considering this outcome as the outcome of the
experiment $\alpha \cdot \beta$. For the piece of wood and the two considered experiments this product experiment consists of
choosing whether we want to test if the piece of wood burns well `or' whether we want to test if the piece of wood floats on water,
one of the two, and then to perform this chosen experiment, and collect the outcome. It is clear that $\alpha \cdot \beta$ will
give with certainty the outcome `yes' if and only if both experiments $\alpha$ `and' $\beta$ will give with certainty the outcome
`yes'. This proves that $\alpha \cdot \beta$ is an experiments that tests whether the system `has' both properties $a$ and $b$ at
once.

At first sight strangely enough, our analysis shows that to test the property $a$ `and' $b$ we need to use the
experiment $\alpha$ `or' $\beta$. The logical `and' changes
into a logical `or' if we shift from the properties to the experiments to tests them. That is the reason that it makes sense to
attribute a lot of properties at once, or the conjunction of all these properties, to an entity, even if these properties cannot be
tested at once. 

If the reader reflects on how we conceive ordinary reality around us, he will find that this is exactly
what we do. We say that a Volvo is a strong car `and' is $x$ meters long, for example, because we know that if we would test one of
the two properties, this one would give us an outcome `yes' with certainty. Although testing the strength of the Volvo, by hitting
it against a wall, would definitely change its length.

\subsection{How Infinity Comes In}
Remark that the definition that we have given for the product experiment works for any number of properties. Suppose that we
have a set of properties $(a_i)_i$ of any size, and for each property an experiment $\alpha_i$ to test it, then the product
experiment $\Pi_i\alpha_i$, consisting of choosing one of the experiments $\alpha_j$ and performing it, and collecting the outcome,
tests all of the properties $a_i$ at once. The `product operation' does not pose any problems with infinity. If it is physically
relevant to consider an infinite set of properties for an entity, then the product experiment tests the conjunction of this infinite
set of properties.

\subsection{What About Probability?}

What we have explained so far is old stuff, and can be found in the earlier articles that have been written on
the subject\cite{Aerts1981,Aerts1982,Piron1976,Piron1989,Piron1990,Aerts1983a,Aerts1983b}. From here on we want to start to
introduce new things, more specifically how probability should be introduced and understood. We have considered only
situations so far where the state of the entity was such that the considered experiments would give the outcome `yes'
with `certainty', and the concept `certainty' has been used somewhat loosely. Would it be possible to introduce the concept
`probability' such that what we have called `certainty' in the foregoing coincides with the notion of probability equal to 1? 

Let
us explain first why this is not a trivial matter, and why it even seems impossible at first sight. We will see in the following
sections that indeed we have to introduce probability in a completely different manner than usually is done to be able to see clear
in the situation and to solve the problem. Consider for now that we have introduced probability in the conventional way. 

This means
that for a state $p$ of the entity and an experiment $\alpha$, we consider a number $\mu(\alpha, p)$ between $0$ and $1$, such that
$\mu(\alpha, p)$ is the probability that if the entity is in state $p$ the experiment $\alpha$ gives the outcome `yes'. This means
that we introduce a function

\bea
\mu: \Sigma \times Q &\rightarrow& [0, 1] \\
(\alpha, p) &\mapsto& \mu(\alpha, p)
\eea
where $\Sigma$ is the set of relevant state of the entity $S$ and $Q$ is the set of relevant yes/no-experiments. As we
said already,
$\mu(\alpha, p)$ is the probability that the experiment $\alpha$ gives the outcome `yes' if the entity is in state $p$. 

For
standard probability theory $\mu(\alpha, p)$ is an element of the unit interval $[0, 1]$ of the set of real numbers $\real$. If we
want to express the situation `the state $p$ of the entity $S$ is such that experiment $\alpha$ would give outcomes `yes' with
certainty', by `$\mu(\alpha, p) = 1$', we have to make certain that our analysis of the product experiments remains valid.

\subsection{Probability of Product Experiments}
Let us consider a set of experiments $(\alpha_i)_i$, and suppose that for a certain state $p$ of the entity $S$, all the
probabilities $\mu(\alpha_i, p)$ are given. Consider now the product experiment $\Pi_i\alpha_i$. What could be the meaning and
definition of $\mu(\Pi_i\alpha_i, p)$? Let us analyze this. 

The product experiments $\Pi_i\alpha_i$ consists of choosing one of
the experiments of the set $(\alpha_i)_i$ and performing this chosen experiment. We can express the `act of choice' between the
different elements of the set $(\alpha_i)_i$ by using the notion of probability itself. Indeed, let us suppose that there is
probability $x_i$ that we would choose the experiment $\alpha_i$. This means that our act of choice is described by a set
$(x_i)_i$ of real numbers, such that $x_j \in [0, 1]$, and such that $\sum_ix_i = 1$. With this act of choice, the
probability of the product experiments, the entity being in state $p$, is given by
\be
\mu(\Pi_i\alpha_i, p) = \sum_ix_i\mu(\alpha_i, p)
\ee
Let us express the situation `the state of the entity is such that the product experiment gives with certainty the
outcome yes' by the formula `$\mu(\Pi_i\alpha_i, p) =1$'. The question is now whether this implies that `the state of the entity
is such that all of the experiments $\alpha_i$ give with certainty the outcome yes'. This means that from $\mu(\Pi_i\alpha_i, p)
=1$ should follow $\mu(\alpha_i, p) = 1$ for all $i$. This is only true if all of the $x_i$ are strictly greater than $0$. Let
us prove this.

\bp \label{prop:productprobability}
Consider a product experiment for which the act of choice is defined by means of a set of real numbers $(x_i)_i \in [0, 1]$ such that
$\sum_ix_i = 1$, and hence $\mu(\Pi_i\alpha_i, p) = \sum_ix_i\mu(\alpha_i, p)$. We have: 
\be
\{\mu(\Pi_i\alpha_i, p) = 1 \Rightarrow \mu(\alpha_j, p) = 1\ \forall j\} \Leftrightarrow \{0 < x_j\ \forall j\}
\ee
\ep
\noindent
Proof: Suppose that $x_j = 0$ then $\sum_ix_i\mu(\alpha_i, p) = \sum_{i \not= j}x_i\mu(\alpha_i, p)$. This means that $
\sum_ix_i\mu(\alpha_i, p) = 1  \Leftrightarrow \sum_{i \not= j}x_i\mu(\alpha_i, p) = 1$. This can however never imply that
$\mu(\alpha_j, p) = 1\ \forall j$, since $\mu(\alpha_j, p)$ can be arbitrary without the sum changing its value. Take now
$0 < x_j\ \forall j$ and suppose that there is a $k$ such that $\mu(\alpha_k, p)
< 1$. We have $\mu(\Pi_i\alpha_i, p) = \sum_ix_i\mu(\alpha_i, p) = x_k\mu(\alpha_k, p) + \sum_{i \not= k}x_i\mu(\alpha_i, p)$.
Since $\mu(\alpha_i, p) \in [0, 1]\ \forall i$ we have $\sum_{i \not= k}x_i\mu(\alpha_i, p) \le \sum_{i \not= k}x_i$. And since
$\sum_ix_i = 1$ we have $\sum_{i \not= k}x_i = 1 - x_k$. From this follows that $\mu(\Pi_i\alpha_i, p) \le x_k\mu(\alpha_k, p) +
(1 - x_k) = 1 - x_k(1 - \mu(\alpha_k))$. Because $\mu(\alpha_k, p) < 1$ we have $0 < 1 - \mu(\alpha_k, p)$, and because $0 < x_k$
also $0 < x_k(1 - \mu(\alpha_k))$. From this follows that $1 - x_k(1 - \mu(\alpha_k)) < 1$. So we have proven that
$\mu(\Pi_i\alpha_i, p) < 1$. \qed

\medskip
\noindent
Suppose that the set of experiments $(\alpha_i)_i$ that we consider for the product experiments $\Pi_i\alpha_i$ is finite, and
that we describe the act of choice corresponding to this product experiment by means of the set of numbers $(x_i)_i$, $x_j \in
[0, 1]$ with $\sum_ix_i = 1$. In this case it would make sense to take out all the $\mu(\alpha_j, p)$ for which $x_j = 0$,
because anyhow the probability to choose such a $\alpha_j$ during the act of choice is $0$, and redefine a new set of experiments
$(\alpha_k)_k$ for the product experiment that only retain those ones for which $0 < x_k$. From proposition
\ref{prop:productprobability} follows that for this new set the property that we want is satisfied: the outcome `yes'
is certain for the product experiment if and only if the outcome `yes' is certain for all the component experiments, where
`certainty' is defined as `probability equals 1'. 

For infinite sets this however does not work in an obvious way. Suppose for
example that we have a set of experiments coordinated by an interval $I$ of the set of real numbers $\real$. Hence $(\alpha_i)_{i
\in I}$ is our set. The probability that one of the $\alpha_i$ is chosen will in general be equal to $0$ then, because the
requirement $\sum_ix_i = 1$ forces the individual $x_i$'s to be $0$. It is common to express everything by means of integrals
instead of sums in this situation. For example $\sum_ix_i = 1$ will be replaced by $\int_I\rho(x)dx = 1$, where $\rho$ is a
function from $I$ to $[0, 1]$. For an interval $[j, k] \subset I$ we interpret $\int_{[j, k]}\rho(x)dx$ as the
probability that an element $\alpha_i$ is chosen with $i \in [j, k]$.

There is an aspect that is very uncomfortable from the
physical point of view with this standard description of the infinite case. The
probability to choose `one' of the $\alpha_i$'s equals $0$, while in reality it are the $\alpha_i$'s individually that are
performed in the laboratory when real tests are executed. So although the probability to choose one of the $\alpha_i$'s equals
$0$, such one of the $\alpha_i$'s `gets' chosen in reality. This problem is known in standard probability theory. One
usually concludes by stating that probability equals $0$ does not mean impossibility. That is also the reason that in this case
probability equals 1 cannot be interpreted as `certainty'.

Because of the fundamental nature of the approach that we are
developing, we cannot be satisfied with this state of affairs. We will show a way to solve this problem, but it demands
the introduction of a new type of probability theory that is a generalization of standard probability theory.

\section{Subset Probability}

In this section we introduce a new type of probability description, where a probability will not be a number of the interval $[0, 1]$,
but a subset of this interval. Let us tell how we arrive at this model.

\subsection{Product Experiments and Subsets} \label{subsec:productexperiment}

We consider again the situation where we have at our disposal a set of yes/no-experiments $(\alpha_i)_i$, and a physical entity $S$
in a state $p \in \Sigma$. In the foregoing sections we have described the act of choice by introducing a set of real numbers $x_i
\in [0, 1]$ with $\sum_ix_i = 1$. Let us reflect a little bit more to see whether this is a good way to describe the act of choice
that we have in mind when we have introduced the idea of a product experiment.

Such a set of numbers $x_i$ means that we consider
the situation as if there is a `fixed' probability for each one of the component experiments $\alpha_i$ to be chosen, and this
fixed probability is represented by the number $x_i$. We have put forward this description because it seems to be the obvious one
starting from standard probability. We can however also look at the situation in a different way. We can consider
the act of choice to be such that we do not have fixed probabilities for each one of the $\alpha_i$'s, but that we can choose each
time again in a different way. A description of this idea would be that we choose each time again among all the possible sets of
numbers
$(x_i)_i \in [0, 1]$, with $\sum_ix_i = 1$, and apply then this specific chosen set for a calculation of the probability of the
product experiment. This means of course that the probability of the product experiment can have different values depending on the
chosen set of numbers $(x_i)_i$ that describe the act of choice. We can prove the following proposition.

\bp \label{prop:productexperiment}
Consider a physical entity $S$ with set of states $\Sigma$ and set of yes/no-experiments $Q$. Suppose that $(\alpha_i)_i$ is a
set of experiments, and $\Pi_i\alpha_i$ the product experiment. If we allow all possible acts of choice described each time by a
set of real numbers $(x_i)_i \in [0, 1]$ such that $\sum_ix_i = 1$, then $\mu(\Pi_i\alpha_i, p)$ can be any number of the
convex hull $V(\{\mu(\alpha_i, p)\})$ of the set $\{\mu(\alpha_i, p)\}$.
\ep
\noindent
Proof: Consider a particular act of choice represented by the set of numbers $(x_i)_i \in [0, 1]$ such that $\sum_ix_i = 1$. We
have
$\mu(\Pi_i\alpha_i, p) = \sum_ix_i\mu(\alpha_i, p) \in V(\{\mu(\alpha_i, p)\})$. On the other hand, suppose that $x \in
V(\{\mu(\alpha_i, p)\})$. This means that there exists a set of numbers $(x_i)_i \in [0, 1]$ with $\sum_ix_i = 1$ such that $x =
\sum_ix_i\mu(\alpha_i, p)$. This means that if we consider the act of choice described by $(x_i)_i$ for $\Pi_i\alpha_i$, then
$\mu(\Pi_i\alpha_i, p) = x$. \qed

\medskip
\noindent
If we consider all possible acts of choice, the convex hull represents the subset of $[0, 1]$ that described the probability
involved. The general situation is however that we do not have to consider all possible acts of choice. Sometimes many different
possible acts of choice are involved, but we lack knowledge about which acts of choice realize during an experiments. This situation
can result again in the probability being represented by one number of the interval $[0, 1]$. This means that to capture the most
general possible situation we should represent the probability by a subset of the interval $[0, 1]$. The subset is the convex hull if
all possible acts of choice are considered in an identifiable way, while the subset is a singleton if there is only one act of
choice, or if all possible acts of choice are unidentifiable, which results again in one act of choice probabilistically distributed
over several `hidden' acts of choice.

\subsection{What About Single Experiments} \label{subsec:singleexperiments}
If we let a subset of $[0, 1]$  correspond with the probability for a product experiment, we seemingly give a special status to the
product experiments as compared to single experiments. If we reflect well however we can see that in fact each experiment is a
product experiment. Always when we consider an experiment we choose the location where to do the experiment, the time when to
perform it, and we also choose among the possible setups, etc \ldots. Of course, these choices are made without taking them into
account because they lead to almost identical situations. In principle however there will be a little subset of the interval $[0, 1]$
centered around one point that in traditional probability theory is taken to represent the probability.

Let us reflect somewhat about the foundations of probability theory itself in this respect. A lot of
different interpretations for probability have been proposed. Much of the discussion was between those defending the
`relative frequency' interpretation and those being in favor of the `subjective' interpretation of probability. Our concern here in
not about this type of issue. In fact, the interpretation of probability is rather obvious in our scheme. 

If we
consider a physical entity $S$ in state $p$, and an experiment $\alpha$, then we consider the probability
$\mu(\alpha, p)$ to be an element of reality, something that `is' there, that expresses the tendency of the experiment $\alpha$ to
give the outcome `yes' if $\alpha$ would be performed. We consider this tendency to be there also when the experiment is not
performed, and in fact -- think of our discussion of property -- specifically to be there when the experiment is not performed.
When the experiment is performed, and repeated very often on the entity prepared in an identical state $p$, the sequence that is
formed by the relative frequency is related to this tendency, in the sense that it allows us to describe this tendency by means of a
number, namely the limit of the sequence of relative frequency. This is our interpretation of standard probability. This means that
there is no contradiction between the subjective and relative frequency interpretation within our approach. 

As we mentioned already, even for a single experiment we always make a choice whenever we perform this single experiment. The
presence of the factor choice will make that different sequences of relative frequency, although they appear from
situations that we classify as repeated experiments, will in general give rise to different limits. These different limits should all
be contained in a `good' description of the probability, certainly if we want the probability to express the `real' presence of a
tendency, because such a tendency should be independent of the not to control variations on the context of the experiments. We want
the probability to express a tendency of the physical entity towards contexts of experiments that we classify as equivalent, although
they are perhaps not equivalent if we would be able to control them better. The situation of the product experiment is in fact a
rough example of this phenomenon that however in principle is always present.

\subsection{Introducing the Subset Probability}
We introduce probability in the following way. Let us consider a physical entity $S$ in a state $p \in \Sigma$ and a
yes/no-experiment $\alpha \in Q$ to be performed. Because of the fluctuations on the context related to the yes/no-experiment
$\alpha$, the sequence of relative frequency will possibly converge to different limits, depending on the `choices' that are made
between the different possible hidden contexts.

\bd [Subset Probability]
Consider a physical entity $S$ with set of states $\Sigma$ and set of yes/no-experiments $Q$. The probability for $\alpha$ to give
outcome `yes' if the entity $S$ is in state $p$ is a function
\bea
\mu: Q \times \Sigma &\rightarrow& {\cal P}([0, 1]) \\
(\alpha, p) &\mapsto& \mu(\alpha, p)
\eea
where ${\cal P}([0, 1])$ is the set of all subsets of $[0, 1]$. The subset $\mu(\alpha, p)$ is the collection of the limits of
relative frequency for outcome `yes' of repeated application of experiment $\alpha$ on the entity $S$ in state $p$. When
$\mu(\alpha, p) =
\emptyset$ this means that the experiment $\alpha$ cannot be performed on the entity $S$ in state $p$. We call $\mu$ a `subset
probability'.
\ed
\noindent
Standard probability theory is retrieved as a special idealized case of our probability theory if all the subsets that
are images of the subset probability are singletons.

\subsection{Inverse Experiments}
We have considered always the probability for a yes/no-experiment to give the outcome `yes'. We introduce the probability for outcome
`no' by means of the inverse experiment.

\bd [Inverse Experiment] \label{def:inverseexperiment}
Consider a physical entity $S$ with set of states $\Sigma$, set of yes/no-experiments $Q$ and subset probability $\mu$. The
inverse experiment
$\widetilde\alpha$ of $\alpha \in Q$ is the same experiment where `yes' and `no' are interchanged. We suppose that if $\alpha \in Q$
then also $\widetilde\alpha \in Q$. 
\ed
\noindent
To be able to express the relation between $\mu(\alpha, p)$ and $\mu(\widetilde\alpha, p)$ we introduce an additional
structure on ${\cal P}([0, 1])$.
\bd
For a subset $V \in {\cal P}([0, 1])$ we define $1 - V = \{x\ \vert x \in [0, 1], 1-x \in V\}$. 
\ed

\bp
For $V, W, (V_i)_i \in {\cal V}$ we have:
\bea
V \subset W &\Rightarrow& 1 - V \subset 1 - W  \label{eq:monotone} \\
1 - \cap_iV_i &=& \cap_i(1 - V_i) \label{eq:tildeintersection} \\
1 - (1 - V) &=& V \label{eq:tildeidempotent} \\
1 - [0, 1] &=& [0, 1] \\
1 -\emptyset &=& \emptyset
\eea
\ep
\noindent
Proof: Suppose that
$V
\subset W$, and consider
$x
\in
1 - V$. This means that
$1-x
\in V$ and hence $1-x \in W$. As a consequence we have $x \in 1 - W$. This proves \ref{eq:monotone}. We have $\cap_iV_i \subset
V_j\
\forall j$. From \ref{eq:monotone} follows then that $1 -\cap_iV_i \subset 1 - V_j \ \forall j$. As a consequence we
have $1 - \cap_iV_i \subset \cap_i(1 -V_i)$. Consider $x \in \cap_i(1 -V_i)$. This means that $x \in
1 - V_j\ \forall j$. Hence $1-x \in V_j\ \forall j$. From this follows that $1-x \in \cap_iV_i$, and hence $x \in
1 - \cap_iV_i$. This shows that $\cap_i(1 - V_i) \subset 1 - \cap_iV_i$. This proves
\ref{eq:tildeintersection}. We have $x \in V \Leftrightarrow 1-x \in 1 - V  \Leftrightarrow 1-(1-x) = x \in
1 - (1 - V))$. This proves \ref{eq:tildeidempotent}. \qed

\bp
Consider a physical entity $S$ with set of states $\Sigma$, set of yes/no-experiments $Q$ and subset probability $\mu$. For $\alpha
\in Q$ and $p \in \Sigma$ we have:
\be
\mu(\widetilde\alpha, p) = 1 - \mu(\alpha, p) \label{eq:tildemorph}
\ee
\ep
\noindent
Proof: Consider $x \in \mu(\widetilde\alpha, p)$. This means that there is a sequence $(x_i)_i$ of relative frequency for outcome
`yes' of $\widetilde\alpha$ such that $\lim x_i \in \mu(\widetilde\alpha, p)$. From definition \ref{def:inverseexperiment} follows
that $(x_i)_i$ is a sequence of relative frequency for outcome `no' for the experiment $\alpha$. Then $(1-x_i)_i$ is a sequence of
relative frequency for outcome `yes' for $\alpha$. If $\lim x_i = x$ then $\lim (1-x_i) = 1-x$. This shows that $1-x \in \mu(\alpha,
p)$, and hence $x \in 1 - \mu(\alpha, p)$. So we have proven that $\mu(\widetilde\alpha,
p) \subset  1 - \mu(\alpha, p)$. Consider $y \in 1 - \mu(\alpha, p)$, then $1-y \in \mu(\alpha, p)$. This means that
there is a sequence of relative frequency $(x_i)_i$ for outcome `yes' for $\alpha$ such that $\lim x_i = 1-y \in \mu(\alpha, p)$.
The sequence of relative frequency for outcome `no' for $\alpha$, and hence the sequence of relative frequency for outcome `yes' for
$\widetilde\alpha$, is then given by $(1-x_i)_i$. This means that $\lim (1-x_i) = 1-(1-y) = y \in \mu(\widetilde\alpha, p)$. So we
have proven that $1 - \mu(\alpha, p) \subset \mu(\widetilde\alpha, p)$. This proves \ref{eq:tildemorph}. \qed

\subsection{Subset Probability and Product Experiments}
In section \ref{subsec:productexperiment} we have analyzed how probability behaves in relation with the product
experiment. In proposition \ref{prop:productexperiment} it is shown that $\mu(\Pi_i\alpha_i, p)$ can be any
number of the convex hull $V(\{\mu(\alpha_i, p)\})$. This is due to the fact that we consider all acts of choice to
be possible identifiable acts of choice. There is a subtle matter involved that we will identify first.

Consider again a set of yes/no-experiments $(\alpha_i)_i$. If we define $\Pi_i\alpha_i$ as the experiment that consists of
choosing one of the $\alpha_i$ and then performing this experiment, we implicitly suppose that we know how to choose
consciously between all of the $\alpha_i$. And if we know how to choose consciously between all of the $\alpha_i$, it also
means that we can identify all of the $\alpha_i$, {\it i.e.} we know exactly which one of the $\alpha_i$ is performed. The set of
all these possible conscious acts of choice where each of the $\alpha_i$ is individually identifiable leads to the convex hull
$V(\{\mu(\alpha_i, p)\}$, as we have shown in section \ref{subsec:productexperiment}. For single experiments we have remarked in
section \ref{subsec:singleexperiments} that there is also always a process of choice involved. It is however always the case that
during the actual experimental process for a single experiment a process of `unconscious' choice takes place. That is one of the
reasons that a single experiment executed repeatedly on a physical entity in the same state can give rise to different outcomes. We
have proven in other work that the indeterminism of quantum mechanics can even be fully explained in this way, an approach that we
have called the hidden measurement
approach \cite{Aerts1986,Aerts1987,Aerts1988,Aerts1993,Aerts1994,AA1997,AACDDV1997,Aerts1998,AS1998,AADL1999,ACS1999,AS2002}.

This means that the
presence of choice does not necessarily lead to the probability being defined on a subset rather than on a singleton, as it
is the case for standard probability theory. If the `choice' is randomized again, because it is an unconscious choice, that
cannot be consciously manipulated, this process of randomization will again lead to a probability defined on a singleton.
The other extreme case is when the process of choice is completely conscious and can be manipulated at will. In this case
the probability will give rise to a convex set. The general situation is however the one that is neither a singleton, which
is one extreme, nor a convex set, which is the other extreme.

For the case of the product experiment $\Pi_i\alpha_i$ of a set of experiments $(\alpha_i)_i$, we will make the hypothesis
that at least each of the experiments individually can be chosen as a single repeated experiment. This means that the
procedure that gives rise to the sequence of relative frequency for outcome `yes', the physical entity being in state $p$, is
the following: one of the experiments $\alpha_j$ is chosen, and this experiment $\alpha_j$ is repeated to give rise to a
sequence of relative frequency. If this is the procedure that we introduce for the product experiment, we can show that the
probability of the product experiment equals the set theoretical union of the probabilities of the single experiments. 

\bd [Product Experiment Procedure] \label{def:productprocedure}
Consider a physical entity $S$ with set of states $\Sigma$, set of yes/no-experiments $Q$ and subset probability $\mu$. For
a set of yes/no-experiments $(\alpha_i)_i \in Q$ we define the procedure of repeated experiments for $\Pi_i\alpha_i$
as follows. One of the experiments $\alpha_j$ is chosen, and this experiment is repeatedly executed on the entity $S$
prepared in the same state $p \in \Sigma$ to give rise to a sequence of relative frequency for outcome `yes' of $\alpha_j$.
This sequence is one of the sequences to define $\mu(\Pi_i\alpha_i, p)$.
\ed

\bp
Consider a physical entity $S$ with set of states $\Sigma$, set of yes/no-experiments $Q$ and subset probability $\mu$. For
a set of yes/no-experiments $(\alpha_i)_i \in Q$, and for $p \in \Sigma$ we have:
\be
\mu(\Pi_i\alpha_i, p) = \cup_i\mu(\alpha_i, p) \label{eq:productprobability}
\ee
\ep
\noindent
Proof: If we execute the procedure for the product experiment $\Pi_i\alpha_i$ as in definition \ref{def:productprocedure} we
find a sequence of relative frequency with a limit that is an element $\mu(\alpha_j, p)$ for one of the $\alpha_j$ chosen
from the set $(\alpha_i)_i$. This proves \ref{eq:productprobability}. \qed

\section{Reality as a Special Case of Probability}
After a large detour into mathematical subtleties related to probability theory, we come back now to the origin of our proposal. We
want to show that with the subset probability we can express consistently the concept of `certainty' as `probability equals
1'. Of course, since probability is now considered to be a subset, the statement `probability equals 1' from standard
probability theory has to be replaced by the statement `probability is a subset of $\{1\}$'.

\subsection{Certainty for Several Experiments}
Let us prove that with the subset probability the certainty for several experiments is expressed well by identifying
certainty with probability contained in the singleton $\{1\}$.

\bd [Certainty]
Consider an entity $S$ with set of states $\Sigma$, set of yes/no-experiments $Q$, equipped with a subset probability $\mu: Q
\times \Sigma \rightarrow {\cal P}([0, 1])$. For $\alpha \in Q$ we say
\be
\alpha\ {\rm gives\ with\ certainty\ yes\ for\ }S\ {\rm in\ state\ } p \Leftrightarrow \mu(\alpha, p) \subset \{1\}
\ee
\ed

\bp \label{prop:certaintyprobability}
Consider an entity $S$ with set of states $\Sigma$, set of yes/no-experiments $Q$, equipped with a subset probability $\mu: Q
\times \Sigma \rightarrow {\cal P}([0, 1])$. For $(\alpha_i)_i \in Q$ and $p \in \Sigma$ we have:
\be
\mu(\Pi_i\alpha_i) \subset \{1\} \Leftrightarrow \mu(\alpha_i, p) \subset \{1\}\ \forall i
\ee
\ep
\noindent
Proof: From \ref{eq:productprobability} we know that $\mu(\Pi_i\alpha_i, p) = \cup_i\mu(\alpha_i, p)$. We have
$\cup_i\mu(\alpha_i, p) \subset \{1\} \Leftrightarrow \mu(\alpha_i, p) \subset \{1\}\ \forall i$. \qed

\subsection{Back to the Original Problem}
Let us come back to what we discussed in section \ref{subsec:real}. We claimed there that we believe that a piece of wood
has both properties $a$ and $b$ at once, the property $a$ of `burning well', and the property $b$ of `floating on water'. This
is because if we would perform one of the tests of these properties, so trying out whether it burns well, which we called test
$\alpha$, or trying out whether it floats, which we called test $\beta$, this would in any case deliver us a positive outcome with
certainty. Introducing the subset probability as we have done in the last few sections makes it possible to replace the concept of
certainty by that of probability contained in the singleton $\{1\}$. Following proposition \ref{prop:certaintyprobability} we indeed
know that
$\mu(\alpha
\cdot \beta, p) \subset \{1\} \Leftrightarrow \mu(\alpha, p) \subset \{1\}$ and $\mu(\beta, p) \subset \{1\}$, because the subset
probability that we have introduced describes well the idea that certainty is independent of whether we choose to perform $\alpha$ or
$\beta$. This is effectively what we have in mind when we introduce certainty speaking about things that really exist.

At first sight one might think that we have perhaps attempted too strong to describe certainty by means of
probability. Indeed, could it not be claimed that the standard probabilistic approach has just the advantage of making the concept
of certainty somewhat weaker and replacing it by probability equals 1, which does not really means certainty, but something like
`very very close to certainty'. And is such a concept of `very very close to certainty' not more realistic than absolute certainty?
We totally agree with this. Often it are cases of `very very close' to certainty that appear in the world around us. In fact, it is
this remark that will make it possible for us to reveal much more profoundly the motivation of all what we are doing here. We do not
want so much to concentrate on how to describe `complete certainty', because indeed such `complete certainty' should be a kind of
extreme case. More profound to our motivation is that the level of certainty, whether this level is the extreme case of complete
certainty, or whether it is very very close to certainty, should be in equilibrium, in the sense that this level should be attained
independent of whether we consider one property or whether we consider several properties at once. This is what standard probability
does not accomplish and what we do accomplish by means of our subset probability. As a bonus it would be nice to have within the
mathematical model a way available to express also the ideal situation of complete certainty. Also in the mathematical model this
should be a extreme case. Also this is not accomplished by standard probability theory and it is by our subset probability. Let us
make this more clear in the next section when we formally consider situations close to certainty. 

\subsection{Situations Close to Certainty} \label{prop:closetocertainty}
Suppose that we consider the situation where we want to express that the outcomes of an experiment $\alpha$ is very close to
certainty. We can do this in an obvious way by demanding that $\mu(\alpha, p) \subset [1-\epsilon,1]$
where $\epsilon$ is a small real number.

\bp
Consider a physical entity $S$ with set of states $\Sigma$ and set of yes/no-experiments $Q$, equipped with a subset probability
$\mu: Q \times \Sigma \rightarrow {\cal P}([0, 1])$. For a set of experiments $(\alpha_i)_i$ and $p \in \Sigma$ we have:
\be
\mu(\Pi_i\alpha_i, p) \subset [1-\epsilon, 1] \Leftrightarrow \mu(\alpha_i, p) \subset [1-\epsilon, 1]\ \forall i
\label{form:equivalenceindivprod}
\ee
\ep
\noindent
Proof: From \ref{eq:productprobability} we know that $\mu(\Pi_i\alpha_i, p) = \cup_i\mu(\alpha_i, p)$. We have
$\cup_i\mu(\alpha_i, p) \subset [1-\epsilon, 1] \Leftrightarrow \mu(\alpha_i, p) \subset [1-\epsilon, 1]\ \forall i$. \qed

\medskip
\noindent
This proposition proves that the level of uncertainty remains the same whether we consider one of the experiments $\alpha_j$ or
whether we consider the product experiment $\Pi_i\alpha_i$. And this should be the case. This is exactly what expresses what we
intuitively have in mind when we think of the procedure of testing several properties at once. Our level of uncertainty should not
depend on the number of properties tested. 

We cannot formulate an equivalent proposition within standard probability theory, because
of the problems that we have mentioned already in the foregoing sections. Indeed, in standard probability theory we are forced to
define the probability related to $\Pi_i\alpha_i$ by means of a convex combination of the probabilities of the $\alpha_i$, for
example expressed by a set of numbers $(x_i)_i$, such that $x_i \in [0, 1]$ and $\sum_ix_i = 1$, and the demand that this convex
combination probability, expressed by $\sum_ix_i\mu(\alpha_i, p)$, is contained in an interval $[1-\epsilon, 1]$ does not imply that
the single probabilities $\mu(\alpha_j, p)$ have to be contained in this same interval. Except when we would demand that
$0 < x_i\ \forall i$ (see proposition \ref{prop:productprobability}). But if we demand $0 < x_i\ \forall i$ we come into problems
with infinite sets of properties and related experiments.

\subsection{Transfer of Other Statements}

It is clear from proposition \ref{prop:closetocertainty} that the equivalence between statements expressed about a collection of
properties, or their tests, and the individual properties, or test, is not limited to statements about closeness to certainty. The
interval $[1 - \epsilon, 1]$ in formula \ref{form:equivalenceindivprod} can be replaced by an arbitrary subset of the interval $[0,
1]$, and remains valid.

\bp
Consider a physical entity $S$ with set of states $\Sigma$ and set of yes/no-experiments $Q$, equipped with a subset probability
$\mu: Q \times \Sigma \rightarrow {\cal P}([0, 1])$. For a set of experiments $(\alpha_i)_i$, $p \in \Sigma$ and $A \subset [0, 1]$ we
have:
\be
\mu(\Pi_i\alpha_i, p) \subset A \Leftrightarrow \mu(\alpha_i, p) \subset A\ \forall i
\ee
\ep
\noindent
Proof: From \ref{eq:productprobability} we know that $\mu(\Pi_i\alpha_i, p) = \cup_i\mu(\alpha_i, p)$. We have
$\cup_i\mu(\alpha_i, p) \subset A \Leftrightarrow \mu(\alpha_i, p) \subset A\ \forall i$. \qed

\medskip
\noindent
So the probability of the product test is contained in an arbitrary subset of $[0, 1]$ iff the probabilities of all the component tests
are contained in this subset. A special case of this is that the probability of the product test is a singleton, and hence
corresponds to one converging series of relative frequency, iff the probability of all the component tests equal this same
singleton and hence also correspond to this one convergent series of relative frequency.

\bp
Consider a physical entity $S$ with set of states $\Sigma$ and set of yes/no-experiments $Q$, equipped with a subset probability
$\mu: Q \times \Sigma \rightarrow {\cal P}([0, 1])$. For a set of experiments $(\alpha_i)_i$, $p \in \Sigma$ and $c \in [0, 1]$ we
have:
\be
\mu(\Pi_i\alpha_i, p) \subset \{c\} \Leftrightarrow \mu(\alpha_i, p) \subset \{c\}\ \forall i
\ee
\ep
\noindent
Proof: From \ref{eq:productprobability} we know that $\mu(\Pi_i\alpha_i, p) = \cup_i\mu(\alpha_i, p)$. We have
$\cup_i\mu(\alpha_i, p) \subset \{c\} \Leftrightarrow \mu(\alpha_i, p) \subset \{c\}\ \forall i$. \qed

\medskip
\noindent
So we see that by means of the subset probability we can not only express the idealized situations of complete certainty, as the
probability contained in singleton
$\{1\}$, but also the idealized situation of a series of relative frequency that converges to one specific limit, as the 
probability contained in an arbitrary singleton.

\section{SEP: The Category of State Experiment Probability Systems} \label{sec:phys}
In the foregoing section we have introduced the subset probability for a physical entity $S$. In this section we introduce the
mathematical structures involved, independent of whether they are used to describe a physical entity. The reason to do so, is that in
this way these structures can be studied independent of their physical meaning. They can then be used as a mathematical model for the
description of a physical entity. This section is self-contained from a mathematical point of view. We introduce
specifically the mathematics that is needed for a mathematical model of the physical situation that we have considered in the
foregoing sections. We introduce immediately the categorical structure.

\subsection{State Experiment Probability Systems}
Let us first introduce the mathematical structure of a state experiment probability system.

\bd [State Experiment Probability System]
A state experiment probability system or SEP $(\Sigma, Q, \Pi,\ \widetilde{ }\ , \mu)$ or shorter $(\Sigma, Q, \mu)$ consists
of two sets
$\Sigma$ and $Q$ and a function
\be
\mu: Q \times \Sigma \rightarrow {\cal P}([0, 1])
\ee
On
$Q$ there exists a product, that associates with each family
$(\alpha_i)_i
\in Q$ an element
$\Pi_i\alpha_i
\in Q$ such that for $p \in \Sigma$
\be
\mu(\Pi_i\alpha_i, p) = \cup_i\mu(\alpha_i, p)
\ee
There
also exists an inverse operation on
$Q$, which is a function
$\widetilde{ }: Q
\rightarrow Q$ such that for $\alpha, (\alpha_i)_i \in Q$ and $p \in \Sigma$ we have:
\bea
\widetilde{\widetilde\alpha} &=& \alpha \\
\widetilde{\Pi\alpha_i} &=& \Pi_i\widetilde\alpha_i \\
\mu(\widetilde\alpha, p) &=& \widetilde{\mu(\alpha, p)}
\eea
where $\widetilde{\mu(\alpha, p)} = \{1-x\ \vert x \in \mu(\alpha, p)\}$. There exists a unit element $\tau \in Q$ such that for $p
\in
\Sigma$:
\be
\mu(\tau, p) = \{1\} \label{eq:unit}
\ee
\ed

\subsection{The Morphisms of State Experiment Probability Systems}
Consider two state experiment probability systems
$(\Sigma, Q, \mu)$ and
$(\Sigma',Q', \mu')$. These
state experiment probability systems respectively describe entities $S$ and $S'$. We will arrive at the notion of a morphism by
analyzing the situation where the entity $S$ is a subentity of the entity $S'$. In that case, the following three natural
requirements should be satisfied:

\smallskip
\noindent i) If the entity $S'$ is in a state $p'$ then the state $m(p')$ of
$S$
is determined. This defines a function $m$ from the set of states of $S'$ to the set of states of $S$; 

\smallskip
\noindent ii) If we consider an experiment $\alpha$ on the entity $S$, then to
$\alpha$
corresponds an experiment $l(a)$ on the ``bigger'' entity $S'$. This defines a function $l$ from
the set of experiments of $S$ to the set of experiments of $S'$. 

\smallskip
\noindent
iii) We want $\alpha$ and $l(\alpha)$ to be two descriptions of the `same' experiment of $S$, once
considered as an entity in itself, once as a subentity of $S'$. In other words we want $\alpha$ and
$l(\alpha)$ to relate to the states $m(p')$ and $p'$ with the same probabilities. This means
that for a state
$p'$ of $S'$ (and a
corresponding state $m(p')$ of $S$) we want the following `covariance principle' to hold:
\be
\mu(\beta, m(p')) = \mu'(l(\beta), p')\ \ \beta \in P, p' \in \Sigma' \\
\ee

\smallskip
\noindent
Furthermore, since the inverse operation is just a switching of `yes' and `no', this switching should have the same result
whether we consider the description of the experiment, namely $\alpha$, on the entity $S$ itself, or the description of the
experiment, namely $l(\alpha)$, on the entity as a subentity. This is expressed by
\be
\widetilde{l(\alpha)} = l(\widetilde\alpha)
\ee
The covariance principle applied for the situation of a set of experiments $(\alpha_i)_i \in Q$ produces in a similar way the
following requirement for the product operation:
\be
l(\Pi_i\alpha_i) = \Pi_il(\alpha_i)
\ee
We are now ready to present a formal definition of a morphism of state experiment systems.

\bd \label{def:morphism}
Consider two state experiment probability systems $(\Sigma, Q, \mu)$ and $(\Sigma',Q', \mu')$. We
say that
\begin{equation}
(m,l):(\Sigma',Q', \mu') \longrightarrow (\Sigma,Q, \mu)
\end{equation}
is a `morphism' (of state experiment probability systems) if $m$ is a function:
\begin{equation}
m: \Sigma' \rightarrow \Sigma
\end{equation}
and $l$ is a function:
\begin{equation}
l: Q \rightarrow Q'
\end{equation}
such that for $\alpha, (\alpha_i)_i \in Q$ and $p' \in \Sigma'$ the following holds:
\bea \label{eq:covar1}
\mu(\alpha, m(p')) &=& \mu'(l(\alpha), p') \label{eq:morphism01} \\
l(\Pi_i\alpha_i) &=& \Pi_il(\alpha_i) \\
\widetilde{l(\alpha)} &=& l(\widetilde\alpha) 
\eea
\ed
\noindent
We introduce the category with objects state experiment probability structures and morphisms the ones that we come to introduce
and denote this category {\bf SEP}.

\section{The Categories SEP and SP} \label{sec:sepandsp}
We can find back the structure of a state property system that we have studied in great detail in earlier articles
\cite{Aerts1999a,ACVVVS1999,Aerts1999b,ACVVVS2001,VanSteirteghem1998,VanSteirteghem2000,VanderVoorde2000,VanderVoorde2000,VanderVoorde2001,ADVV2001,ADVV2002,AD2002}
related to a state experiment probability structure, and describing the properties that are operationally defined by the
experiments of the state experiment probability system. Also the category of state property systems and its morphism, called
{\bf SP}, appears as a substructure of the category {\bf SEP}.

\subsection{The Related State Property System}
Let us repeat the definition of a state
property system.

\bd [State Property System]
We say that $(\Sigma, <, {\cal L}, <, \wedge, \vee, \xi)$ is a state property system if $\Sigma, <$ is a pre-ordered set, ${\cal L}, <,
\wedge, \vee$ is a complete lattice, and $\xi$ is a function:
\be
\xi: \Sigma \rightarrow {\cal P}({\cal L})
\ee 
such that for $p \in \Sigma$, $I$ the maximal element of ${\cal L}$, and $0$ the minimal element of ${\cal L}$, we have:
\bea
I &\in& \xi(p) \label{eq:statprop01} \\
0 &\not\in& \xi(p) \label{eq:statprop02} \\
a_i \in \xi(p)\ \forall i &\Leftrightarrow& \wedge_ia_i \in \xi(p) \label{eq:statprop03}
\eea
and for $p, q \in \Sigma$ and $a, b \in {\cal L}$ we have:
\bea
p < q &\Leftrightarrow& \xi(q) \subset \xi(p) \label{eq:statprop04} \\
a < b &\Leftrightarrow& \forall r \in \Sigma: a \in \xi(r)\ {\rm then}\ b \in \xi(r) \label{eq:statprop05}
\eea
\ed

\bp
Consider a state experiment probability system $(\Sigma, Q, \mu)$. For $p, q \in \Sigma$ and $\alpha, \beta \in Q$ we define:
\bea
p < q &\Leftrightarrow& \forall \gamma \in Q: \mu(\gamma, q) \subset \{1\}\ {\rm then}\ \mu(\gamma, p) \subset \{1\}
\label{eq:preorderstates} \\
\alpha < \beta &\Leftrightarrow& \forall r \in \Sigma: \mu(\alpha, r) \subset \{1\}\ {\rm then}\ \mu(\beta, r) \subset \{1\}
\label{eq:preorderexperiments}\\
\alpha \approx \beta &\Leftrightarrow& \alpha < \beta\ {\rm and}\ \beta < \alpha \label{eq:equivalence}
\eea
then $\Sigma, <$ and $Q, <$ are pre-ordered sets, and $\approx$ is an equivalence relation on $Q$.
\ep
\noindent
Proof: Straightforward verification. \qed

\bd \label{def:propertylattice}
Consider a state experiment probability system $(\Sigma, Q, \mu)$. We define ${\cal L}$ to be the set of equivalence classes of $Q$
for the equivalence relation $\approx$ defined in \ref{eq:equivalence}, and for $a, b \in {\cal L}$ we define $a < b$ iff $\alpha \in
a$ and $\beta \in b$ such that $\alpha < \beta$. We denote $I \in {\cal L}$ the equivalence class of $\tau$ as defined in
\ref{eq:unit}.
\ed

\bp
${\cal L}$ of definition \ref{def:propertylattice} is a complete lattice for the partial order relation $<$, where the infimum of a
set of elements $(a_i)_i \in {\cal L}$ is given by the equivalence class of $\Pi_i\alpha_i$, with $\alpha_j \in a_j\ \forall j$. We
denote this infimum of the set $(a_i)_i$ by $\wedge_ia_i$. The minimal element of ${\cal L}$ is given by $0 = \wedge_{a \in {\cal
L}}a$, $I$ is the maximal element of ${\cal L}$, and the supremum of a set of elements $(a_i)_i$ is given by $\vee_ia_i = \wedge_{a_j <
b\forall j, b\in {\cal L}}b$.
\ep
\noindent
Proof: Since $<$ is a pre-order relation on $Q$, we have that $<$ is a partial order relation on ${\cal L}$. Let us prove that
$\wedge_ia_i$ is an infimum for the set $(a_i)_i$. First we want to show that $\wedge_ia_i$ is a lower bound, hence that $\wedge_ia_i
< a_j\ \forall j$. We know that $\wedge_ia_i$ is the equivalence class of $\Pi_i\alpha_i$ for $\alpha_j \in a_j\ \forall j$. Consider
an arbitrary element $\gamma \in Q$ such that $\gamma \in \wedge_ia_i$, and $r \in \Sigma$ such that $\mu(\gamma, r) \subset \{1\}$.
Since $\gamma \approx \Pi_i\alpha_i$ and hence $\gamma < \Pi_i\alpha_i$, from \ref{eq:preorderexperiments} follows that
$\mu(\Pi_i\alpha_i, r) \subset \{1\}$. We know that $\mu(\Pi_i\alpha_i, r) = \cup_i\mu(\alpha_i, r)$, and hence
we have $\cup_i\mu(\alpha_i, r) \subset \{1\}$. From this follows that $\mu(\alpha_j, r) \subset \{1\}\ \forall j$. Using again
\ref{eq:preorderexperiments} this shows that $\gamma < \alpha_j\ \forall j$. Since $\gamma \in \wedge_ia_i$ and $\alpha_j \in a_j\
\forall j$ we have proven that $\wedge_ia_i < a_j\ \forall j$. Let us prove now that $\wedge_ia_i$ is the greatest lower bound of the
set $(a_i)_i$. Consider an arbitrary $b \in {\cal L}$ such that $b$ is a lower bound of the set $(a_i)_i$. We have
to show that $b < \wedge_ia_i$. Take $\gamma \in b$. Since $b$ is a lower bound we have $\gamma < \alpha_j\ \forall j$. Consider $r
\in \Sigma$ such that $\mu(\gamma, r) \subset \{1\}$. From \ref{eq:preorderexperiments} follows that $\mu(\alpha_j, r) \subset \{1\}\
\forall j$. Then $\cup_i\mu(\alpha_i, r) = \mu(\Pi_i\alpha_i,p) \subset \{1\}$. Again using \ref{eq:preorderexperiments} we have shown
that $\gamma < \Pi_i\alpha_i$. And this proves that $b < \wedge_ia_i$. For any partially ordered set ${\cal L}$ where the infimum of
any set of elements exists, the minimal element is given by $\wedge_{a \in {\cal L}}a$. Let us show that $I$ is the maximal element of
${\cal L}$. Consider any $a \in {\cal L}$, and $\gamma \in a$. Since $\mu(\tau, p) \subset \{1\}\ \forall p \in \Sigma$ we have
$\gamma < \tau$. This proves that $a < I$. For any partially ordered set ${\cal L}$ with a maximal element and where each set $(a_i)_i$
has an infimum, each set $(a_i)_i$ has also a supremum that is given by $\vee_ia_i = \wedge_{a_j <
b\forall j, b\in {\cal L}}b$. This proves that ${\cal L}$ is a complete lattice. \qed

\bd \label{def:xi}
Consider a state experiment probability system $(\Sigma, Q, \mu)$ and the complete lattice ${\cal L}$ of definition
\ref{def:propertylattice}. For $p \in \Sigma$ we define:
\bea
\xi: \Sigma &\rightarrow& {\cal P}({\cal L}) \\
p &\mapsto& \xi(p)
\eea
where
\be
\xi(p) = \{a\ \vert a \in {\cal L}, \mu(\alpha, p) \subset \{1\}, \alpha \in a\}
\ee
\ed
\bp \label{prop:relatedstatprop}
Consider a state experiment probability system $(\Sigma, Q, \mu)$. Consider $\Sigma, <$, with $<$ defined as in
\ref{eq:preorderstates}, ${\cal L}, <, \wedge, \vee$, as in definition \ref{def:propertylattice}, and $\xi$ as in definition
\ref{def:xi}. Then $(\Sigma, <, {\cal L}, <, \wedge, \vee, \xi)$ is a state property system. 
\ep
\noindent
Proof: Consider $p \in \Sigma$, then $\mu(\tau, p) \subset \{1\}$. Hence $I \in \xi(p)$, which proves \ref{eq:statprop01}. Let us
consider $\widetilde\tau \in Q$. We have $\mu(\widetilde\tau, p) \subset \{0\}\ \forall p \in \Sigma$. Let us denote the equivalence
class of $\widetilde\tau$ by $a$. Then $0 < a$. Take an arbitrary element $\gamma \in Q$ such that $\gamma \in 0$. The following is
satisfied: $\forall r \in \Sigma$ such that $\mu(\widetilde\tau, r) \subset \{1\}$ we have $\mu(\gamma, r) \subset \{1\}$. This shows
that $a < 0$, and hence $a = 0$. As a consequence $\widetilde\tau \in 0$. Since $\mu(\widetilde\tau, p) \subset \{0\}\ \forall p \in
\Sigma$ we have $0 \not\in \xi(p)\ \forall p \in \Sigma$, which proves \ref{eq:statprop02}. Consider $(a_i)_i \in {\cal L}$ and
suppose that $\wedge_ia_i \in \xi(p)$. This means that for $\alpha_i \in a_i$ we have $\mu(\Pi_i\alpha_i, p) = \cup_i\mu(\alpha_i, p)
\subset \{1\}$. Hence $\mu(\alpha_j, p) \subset \{1\}\ \forall j$, which proves that $a_j \in \xi(p)\ \forall j$, which proves one of
the implications of \ref{eq:statprop03}. Consider again $(a_i)_i \in {\cal L}$ such that $a_j \in \xi(p)\ \forall j$. With $\alpha_i
\in a_i$ it follows that $\mu(\alpha_j, p) \subset \{1\}\ \forall j$. Hence $\cup_i\mu(\alpha_i, p) = \mu(\Pi_i\alpha_i, p) \subset
\{1\}$. This shows that $\wedge_ia_i \in \xi(p)$ which proves that other implication of \ref{eq:statprop03}. Consider $p, q \in
\Sigma$, such that $p < q$ and consider $a \in \xi(q)$. This means that for $\alpha \in a$ we have $\mu(\alpha, q) \subset \{1\}$.
From \ref{eq:preorderstates} and $p < q$ follows that $\mu(\alpha, p) \subset \{1\}$, and hence $a \in \xi(p)$. This shows that
$\xi(q) \subset \xi(p)$, and so we have proven one of the implications of \ref{eq:statprop04}. Suppose now that $\xi(q) \subset
\xi(p)$, and consider $\gamma \in Q$ such that $\mu(\gamma, q) \subset \{1\}$. If $a$ is the equivalence class of $\gamma$ we have $a
\in \xi(q)$, but then also $a \in \xi(p)$. As a consequence we have $\mu(\gamma, p) \subset \{1\}$. Using \ref{eq:preorderstates} we
have shown that $p < q$. This proves that other implication of \ref{eq:statprop04}. Consider $a, b \in {\cal L}$ such that $a < b$,
and $r \in \Sigma$ such that $a \in \xi(r)$. Take $\alpha \in a$ and $\beta \in b$. We have $\mu(\alpha, r) \subset \{1\}$, and since
$a < b$ it follows from \ref{eq:preorderexperiments} that $\mu(\beta, r) \subset \{1\}$. As a consequence we have $b \in \xi(r)$. This
proves one of the implications of \ref{eq:statprop05}. Suppose that $\forall r \in \Sigma$ we have that $a \in \xi(r)$ implies that $b
\in \xi(r)$. Take $\alpha \in a$, $\beta \in b$ and consider $r \in \Sigma$ such that $\mu(\alpha, r) \subset \{1\}$. This means that
$a \in \xi(r)$, and hence $b \in \xi(r)$. As a consequence we have $\mu(\beta, r) \subset \{1\}$. So we have shown that $\alpha <
\beta$ and hence $a < b$. This proves that other implication of \ref{eq:statprop05}. \qed

\medskip
\noindent
\bd [Related State Property System]
We call $(\Sigma, {\cal L}, \xi)$, as defined in proposition \ref{prop:relatedstatprop}, the state property system related to
the state experiment probability system
$(\Sigma, Q,
\mu)$.
\ed

\subsection{The Related Morphisms}

We will prove that a morphism of {\bf SEP} gives rise to a morphism of {\bf SP}, the category of state property systems as
defined in \cite{Aerts1999a,ACVVVS1999}. Let us put forward the definition of a morphism for a state property system as in
definition 11 of \cite{ACVVVS1999}.

\bd [Morphism of State Property Systems]
Consider two state property systems $(\Sigma, {\cal L}, \xi)$ and $(\Sigma', {\cal L}', \xi')$. We say that
\be
(m, n): (\Sigma', {\cal L}', \xi') \rightarrow (\Sigma, {\cal L}, \xi)
\ee
is a morphism of state property systems, if $m$ is a function:
\be
m: \Sigma' \rightarrow \Sigma
\ee
and $n$ is a function:
\be
n: {\cal L} \rightarrow {\cal L}'
\ee
such that for $a \in {\cal L}$ and $p' \in \Sigma'$ the following holds:
\be
a \in \xi(m(p')) \Leftrightarrow n(a) \in \xi'(p') \label{eq:morphismstatprop}
\ee
\ed

\bp \label{prop:relatedmorphism}
Consider two state experiment probability systems $(\Sigma, Q, \mu)$ and $(\Sigma', Q', \mu')$, and the corresponding state
property systems $(\Sigma, {\cal L}, \xi)$ and $(\Sigma', {\cal L}', \xi')$. For a morphism $(m, l)$ between $(\Sigma, Q, \mu)$
and $(\Sigma', Q', \mu')$ we keep $m$ and define $n: {\cal L} \rightarrow {\cal L}'$ such that $n(a)$ is the equivalence class
in $Q'$ of $l(\alpha)$ with $\alpha \in a$. The couple $(m, n)$ is a morphism between the state property systems
$(\Sigma, {\cal L}, \xi)$ and $(\Sigma', {\cal L}', \xi')$.
\ep
\noindent
Proof: For $p' \in \Sigma'$ consider $a \in \xi(m(p'))$ and $\alpha \in a$. This means that $\mu(\alpha, m(p')) \subset \{1\}$.
Form \ref{eq:morphism01} we have $\mu(\alpha, m(p')) = \mu'(l(\alpha), p')$, and hence $\mu'(l(\alpha), p') \subset \{1\}$. This
shows that $n(a) \in \xi'(p')$ because it is the equivalence class of $l(\alpha)$. This proves one of the implications of
\ref{eq:morphismstatprop}. Suppose that $n(a) \in \xi'(p')$ for $p' \in \Sigma$. This means that the equivalence class of
$l(\alpha)$ for $\alpha \in a$ belongs to $\xi'(p')$. Hence $\mu'(l(\alpha, p')) \subset \{1\}$. Again using \ref{eq:morphism01}
this implies that $\mu(\alpha, m(p')) \subset \{1\}$, and hence $a \in \xi(m(p'))$, which proves the other implication of
\ref{eq:morphismstatprop}. \qed

\bd [Related Morphism]
Let us consider the state property system $(\Sigma, {\cal L}, \xi)$ related to the state experiment probability system
$(\Sigma, Q, \mu)$. We call
$(m, n)$, as defined in proposition
\ref{prop:relatedmorphism}, the morphism of $(\Sigma, {\cal L}, \xi)$ related to the morphism $(m, l)$ of $(\Sigma, Q, \mu)$.
\ed
\noindent
It is easy to see that for any subset $A \in {\cal P}([0, 1])$ we can link a state property system with the state experiment
probability system that we consider. Indeed, for the propositions that we have proved in section \ref{sec:sepandsp} and the
definitions that are given there, we can readily replace the singleton $\{1\}$ by the subset $A$ of the interval $[0, 1]$, and
everything can be proven and defined in an analogous way. Of course, it is the state property system and the morphisms that we
introduced in the foregoing section that describe the state and properties of the physical entity under study. For a further
elaboration of the structure of the state experiment probability system and its relation with the different state property
systems that we can define in this way, it would be interesting to study also these other state property systems, and how they
are interrelated. We will leave this however for future research.

We end this article with a last remark.  Relative
frequencies are rational numbers. This means that if we consider limits of sequences of relative frequency, it does not have to be a
priori so that we have to consider these limits within the completion of $\real$ of $\rational$. There is, to the best of our
knowledge, no operational reason why the completion $\real$  of $\rational$ would be superior. Other completions of
$\rational$ might be a better choice \cite{khrennikov1999}. There are even operational reasons to believe that other completions of
$\rational$ might indeed be a better choice. If we think of the completion used in nonstandard analysis, where infinitesimals are
possible, it would be possible there to explore the problem that we have considered here. Indeed, it might well be possible that
infinite convex of numbers between 0 and 1 can be introduced there, such that all the elements of the convex combination can be
taken to be different from zero while the sum itself remains in the interval $[0, 1]$. This would be another possibility to explore
that could lead to a solution of the problem of how to express `certainty' by means of probability. Since the subset
probability works on the level of ${\cal P}([0, 1])$ and not on the level of $[0, 1]$ itself, it might even be that the structure
that we have investigated in this article can more readily be transposed to the situation where another completion of $\rational$
than $\real$ is considered.

\end{document}